\documentclass[aps,prd,preprintnumbers,nofootinbib,preprint]{revtex4}
\usepackage{bm}
\usepackage{latexsym}
\usepackage{dcolumn}
\usepackage{amsmath,amsfonts,amssymb}
\usepackage{graphicx,epsfig}
\usepackage{amsthm}
\usepackage{url,hyperref}

\usepackage{slashed}

\newcommand{\be}{\begin{eqnarray}}
\newcommand{\ee}{\end{eqnarray}}
\newcommand{\bea}{\begin{eqnarray}}
\newcommand{\eea}{\end{eqnarray}}

\def\comment#1{}

\usepackage{color}

\definecolor{darkred}{rgb}{.8,0,0}

\definecolor{darkblue}{rgb}{0,0,.7}

\definecolor{darkgreen}{rgb}{0,.7,0}

\begin{document}

%
%
\title{Joule-Thomson Expansion of Hayward-AdS black hole}
%
%
%
%
%
\author{Sen Guo$^{1}$} \email[email:~]{quantumguosenedu@163.com}
\author{Jin Pu$^{1,2}$}\email[email:~]{pujin@cwnu.edu.cn}
\author{Qing-Quan Jiang$^{1}$}\email[email:~]{qqjiangphys@yeah.net}
\affiliation{$^1$College of Physics and Space Science, China West Normal University, Nanchong 637002, China\vspace{1ex}}
\affiliation{$^2$School of Physics, University of Electronic Science and Technology of China, Chengdu 610054, China\vspace{1ex}}

%
%
%
%
%
\begin{abstract}
%
%
%
%
%
%
\par\noindent
In this paper, we study Joule-Thomson expansion for Hayward-AdS black hole in the extended phase space, and obtain a Joule-Thomson expansion formula for the black hole. We plot the inversion and isenthalpic curves in the $T-P$ plane, and determine the cooling-heating regions. The intersection points of the isenthalpic and inversion curves are exactly the inversion points discriminating the heating process from the cooling one.\\
\end{abstract}
%
%
%
%
\maketitle
%
%
%
%
%
%
\section{Introduction}
\label{intro}
%
%
\par\noindent
It is well known that black holes as thermodynamic systems have many interesting research fields in theoretical physics. The thermodynamic properties of black holes have been extensively investigated since the first studies of Bekenstein and Hawking \cite{1,2,3,4,5,6}. It sets deep and fundamental connections between the laws of classical general relativity, thermodynamics and quantum mechanics. Since it has a key feature to understand quantum gravity, much attention has been paid to this topic. Black holes as thermodynamic systems have many exciting similarities with general thermodynamic systems. These similarities become more obvious and precise for black holes in AdS space. The research on AdS black hole thermodynamics began with pioneering paper of Hawking and Page, and they found a phase transition between Schwarzschild-AdS black hole and thermal AdS space \cite{7}. Furthermore, thermodynamic properties of charged AdS black holes were studied in \cite{8,9}, and it shown that charged AdS black holes have a Van der Waals like phase transition.
\par
Recently, black hole thermodynamics in AdS space has been intensively studied in the extended phase space where cosmological constant is considered as thermodynamic pressure
\be
P=-\frac{\Lambda}{8\pi},
\label{1-1}
\ee
and its conjugate quantity as thermodynamic volume
\be
V={(\frac{\partial M}{\partial P})}_{S,Q,J},
\label{1-2}
\ee
which can enrich AdS black hole thermodynamics \cite{10}. In the extended phase space (including $P$ and $V$ terms in the first law of black hole thermodynamics), phase transition of charged AdS black hole is similar with that of Van der Waals liquid-gas \cite{11}, and the black hole shares the same $P-V$ diagram and critical exponent with Van der Waals system \cite{12,13,14,15,16,17,18,19,20,21,22,23,24,25,26,27,28,29,30}. In \cite{31,32,33,34}, one can also find the black hole add the second and third-order phase transitions in addition to phase transition of Van der Waals fluid.
\par
Apart from phase transition and critical phenomena, thermodynamic analogies between AdS black hole and Van der Waals system have been creatively generalized to the well-known Joule-Thomson expansion process. Joule-Thomson expansion in thermodynamics is gas at a high pressure passing through a porous plug to a section with a low pressure, and during the expansion, enthalpy is constant. In \cite{35}, \"{O}zg\"{u}r \"{O}kc\"{u} and Ekrem Aydmer have first investigated Joule-Thomson expansion of charged RN-AdS black hole, where the inversion and isenthalpic curves were obtained and the heating-cooling regions were shown in the $T-P$ plane. Subsequently, Joule-Thomson expansions in virous black holes such as the quintessence charged AdS black hole \cite{36}, Kerr-AdS black hole \cite{37}, d-dimensional charged AdS black holes \cite{38}, holographic super-fluids \cite{39}, charged AdS black hole in $f(r)$ gravity \cite{40}, AdS black hole with a global monopole \cite{41}, AdS black holes in Lovelock gravity \cite{42}, charged Gauss-Bonnet black holes \cite{43}, Ads black hole in Einstien-Maxwell-axions theory and AdS black hole in massive gravity\cite{44}, have been extensively investigated.
\par
On the other hand, it is well-known that Bardeen has first studied the black hole with no singularities in the center\cite{45}. Then many regular black holes has been found, such as Hayward black hole \cite{46} and improved Hayward black hole \cite{47,48}, etc. Meanwhile, thermodynamic properties of these regular black holes have been also discussed in \cite{49,50,51,52,53}. In particular, many interesting properties of thermodynamics have been found in Hayward-AdS black hole \cite{54,55,56}. However, Joule-Thomson expansion of Hayward-AdS black hole in the extended phase space has not been studied so far. The main purpose of this paper is to investigate Joule-Thomson expansion of Hayward-AdS black hole.
\par
The remainder of this paper is organized as follows. In Sec.\ref{sec2}, we review thermodynamic properties of Hayward-AdS black hole in the extended phase space. In Sec.\ref{sec3}, by applying Joule-Thomson expansion of the black hole, we obtain the isenthalpic and inversion curves in the $T-P$ plane, and determine the cooling-heating regions. Sec.\ref{sec4} ends up with some conclusions.
%
%
%
\section{The thermodynamic properties of Hayward-AdS black hole}
\label{sec2}
%
%
%
\par\noindent
In this section, we briefly review thermodynamic properties of Hayward-AdS black hole in the extended phase space. The line element of Hayward-AdS black hole is given by \cite{52,57}
\bea
&&ds^{2}=-f dt^{2}+\frac{dr^{2}}{f}+r^{2}d\Omega^{2},\\
&&A=Q_{m}\cos\theta d\phi,\\
&&f(r)=\frac{r^{2}}{l^{2}}+1-\frac{2\mathcal{M}r^{2}}{r^{3}+q^{3}},\label{2-1}
\eea
where $q$ is an integration constant with respect to magnetic charge. $\mathcal{M}$ denotes AMD mass \cite{58,59}, which consists of two parts, the self-interactions of graviton and the nonlinear interactions between graviton and (nonlinear) photon (see \cite{57} for details). The form of AMD mass is expressed as
\be
\mathcal{M}=M+\sigma^{-1}q^{3},
\label{2-2}
\ee
where the parameter $\sigma$ associated with the nonlinear electromagnetic field is viewed as a dynamic variable. $M$ is associated with the condensate of massless graviton, originating from its self-interactions. In neutral limit, the solution reduces to Schwarzschild AdS black hole. Hence, $M$ is viewed as Schwarzschild mass. It is obvious that for any non-zero of $M$, the metric behaves singular at the origin and the existence of the singularity is inevitable. For the Hayward-AdS black hole, the form of the AMD mass is now given by
\be
\mathcal{M}=\sigma^{-1}q^{3},
\label{2-3}
\ee
where the charge term $\sigma^{-1}q^{3}$ is derived from the nonlinear interactions between graviton and (nonlinear) photon.
\par
Thus, the temperature of the black hole is obtained by the first derivative of $f(r)$ at the horizon, i.e.
\be
T=\frac{r_0(r_0^{3}+q^3)^2-2\mathcal{M}r_0l^2q^3+\mathcal{M}r_0l^4}{l^2(q^3+r^3_0)^2},
\label{2-4}
\ee
and the entropy of the black hole is given by
\be
S=\pi r^2_0,
\label{2-5}
\ee
where $r_{0}$ is the horizon radius defined by the largest root of the equation $f(r_{0})=0$. The magnetic charge and its conjugate potential are respectively expressed as
\bea
&&Q_{m}=\frac{q^{2}}{\sqrt{2 \sigma}},  \\
&&\Psi=\frac{3 q^{4}(2{r_{0}}^{3}+q^{3})}{\sqrt{2 \sigma}({q^{3}+{r_{0}}^{3}})^{2}}.
\label{2-6}
\eea
In the extended phase space, cosmological constant and parameter $\sigma$ of the nonlinear electromagnetic field are viewed as thermodynamic variables, so pressure and thermodynamic volume are respectively defined as usual \cite{31,60}
\bea
&&P=-\frac{\Lambda}{8 \pi}=\frac{3}{8\pi l^{2}},\\\label{2-7}
&&V=\frac{4 \pi {r_{0}}^{3}}{3}.
\eea
The first law of thermodynamics and Smarr formula for the black hole are respectively given by
\bea
&&d\mathcal{M}=TdS+VdP+\Psi dQ_{m}+\Pi d\sigma,\\
&&\mathcal{M}=2TS+\Psi Q_{m}-2VP+2\Pi \sigma,
\label{2-8}
\eea
where
\be
\Pi=\frac{q^{6}(2{r_{0}}^{3}-q^{3})}{4 \sigma^{2}({q^{3}+{r_{0}}^{3})}^{2}}.
\label{2-9}
\ee
Here, the mass of the black hole $\mathcal{M}$ should be interpreted as the enthalpy $H$ \cite{12}. And Eq.(14) can be rewritten as
\be
dH=TdS+VdP+\Psi dQ_{m}+\Pi d\sigma.
\label{2-10}
\ee
Then, from Eq.(\ref{2-4}) and Eq.(12), the state equation can been obtained as
\be
P=\frac{T}{2 r_{0}}+\frac{1}{8 \pi {r_{0}}^{2}}+\frac{q^{3}T}{2{r_{0}}^{4}}+\frac{q^{3}}{4 \pi {r_{0}}^{5}}.
\label{2-11}
\ee
In this section, we obtain some thermodynamic quantities of Hayward-AdS black hole. In the next section, we will use these quantities to investigate Joule-Thomson expansions for Hayward-AdS black hole.
%
%
%
\section{Joule-Thomson Expansion}
\label{sec3}
%
%
%
\par\noindent
In this section, we investigate Joule-Thomson expansion for Hayward-AdS black hole. This expansion is characterized by temperature change with respect to pressure. During expansion process of black hole system, the enthalpy remains constant. Thus, Joule-Thomson expansion is an isenthalpy process in the extended phase space. Joule-Thomson coefficient $\mu$, which characterizes the expansion, is given by \cite{61}
\begin{eqnarray}
\mu={(\frac{\partial T}{\partial P})}_{H}.
\label{3-1}
\end{eqnarray}
The cooling-heating regions can be determined by sign of Eq.(\ref{3-1}). Change of pressure is negative since pressure always decreases during  expansion. Temperature may decrease or increase during process. Therefore temperature determines sign of $\mu$. If $\mu$ is negative(positive), heating (cooling) occurs, and so gas warms(cools).
\par
Similar to Joule-Thomson process with fixed particle number for Van der Waals gases, we should consider canonical ensemble with fixed change $q$. In Ref.\cite{35}, Joule-Thomson coefficient is given by
\be
\mu={(\frac{\partial T}{\partial P})}_\mathcal{M}=\frac{1}{C_p}[T({\frac{\partial V}{\partial T}})_p-V].
\label{3-2}
\ee
Setting $\mu=0$, we can obtain the inversion temperature, i.e.
\be
T_{i}=V {(\frac{\partial T}{\partial V})}_{P}.
\label{3-3}
\ee
Here, we use another method to derive Joule-Thomson coefficient $\mu$. From Eqs.(\ref{2-1}) and (\ref{2-7}), pressure $P$ can be rewritten as a  function of $\mathcal{M}$ and $r_0$, i.e.
\be
P(\mathcal{M},{r}_{0})=\frac{6\mathcal{M}{r_{0}}^{2}-3q^{3}-3{r_{0}}^{3}}{8 \pi {r_{0}}^{2}(q^{3}+{r_{0}}^{3})},
\label{3-4}
\ee
then substituting $P(\mathcal{M},r_0)$ into the temperature (\ref{2-4}) yields
\be
T(\mathcal{M},{r}_{0})=\frac{-{r_{0}}^{6}+3\mathcal{M}{r_{0}}^{5}-2q^{3}{r_{0}}^{3}-q^{6}}{2\pi r_{0}{(q^{3}+{r_{0}}^{3})}^{2}}.
\label{3-5}
\ee
Thus, Joule-Thomson coefficient is obtained as
\bea
\mu&=&{\big(\frac{\partial T}{\partial r_{0}}\big)}_\mathcal{M}/{(\frac{\partial P}{\partial r_{0}})}_\mathcal{M} \\ \nonumber
&=&\frac{4r_{0}[q^{6}+8q^{3}({r_{0}}^{3}+2P \pi{r_{0}}^{5})-2({r_{0}}^{6}+4P \pi {r_{0}}^{8})]}{3(q^{3}+{r_{0}}^{3})(2q^{3}-{r_{0}}^{3}-8 P \pi{r_{0}}^{5})}.
\label{3-6}
\eea
Setting $\mu=0$, we get
\be
q^{6}+8q^{3}({r_{0}}^{3}+2P \pi{r_{0}}^{5})-2({r_{0}}^{6}+4P \pi {r_{0}}^{8})=0.
\label{3-7}
\ee
In order to investigate minimum inversion temperature, we set the inversion pressure $P_i=0$ for Eq.(\ref{3-7}). The positive and real root is given by
\be
r_0=\sqrt[3]{(2+\frac{3\sqrt{2}}{2})q^{3}}.
\label{3-8}
\ee
Substituting this root (\ref{3-8}) into Eq.(\ref{2-4}), the inversion temperature is given by
\be
T_{i}=\frac{3\sqrt[6]{2}+4\sqrt[\frac{3}{5}]{4+3\sqrt{2}}P_{i} \pi q^{2}}{6(2+\sqrt{2})\sqrt[3]{8+6\sqrt{2}}\pi q}.
\label{3-9}
\ee
When the inversion pressure $P_i$ is zero, minimum inversion temperature is given by
\be
{T_{i}}^{min}=\frac{1}{\sqrt[\frac{6}{5}]{2}(2+\sqrt{2})\sqrt[3]{8+6\sqrt{2}}\pi q}.
\label{3-10}
\ee
\par\noindent
\begin{figure}
\centering\includegraphics[width=0.4\textwidth]{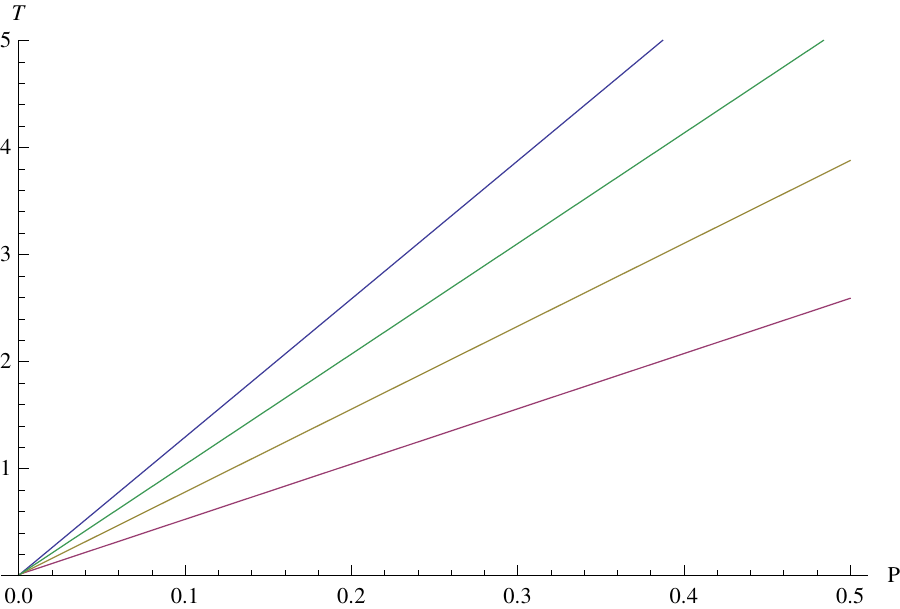}
\caption{Inversion curves of Hayward-AdS black hole in $T-P$ plane. From bottom to top, the curves correspond to $q=2,3,4,5$.}
\label{fig1}
\end{figure}
\par
\begin{figure}
\centering
\label{fig2.a}
\includegraphics[width=4cm,height=3cm]{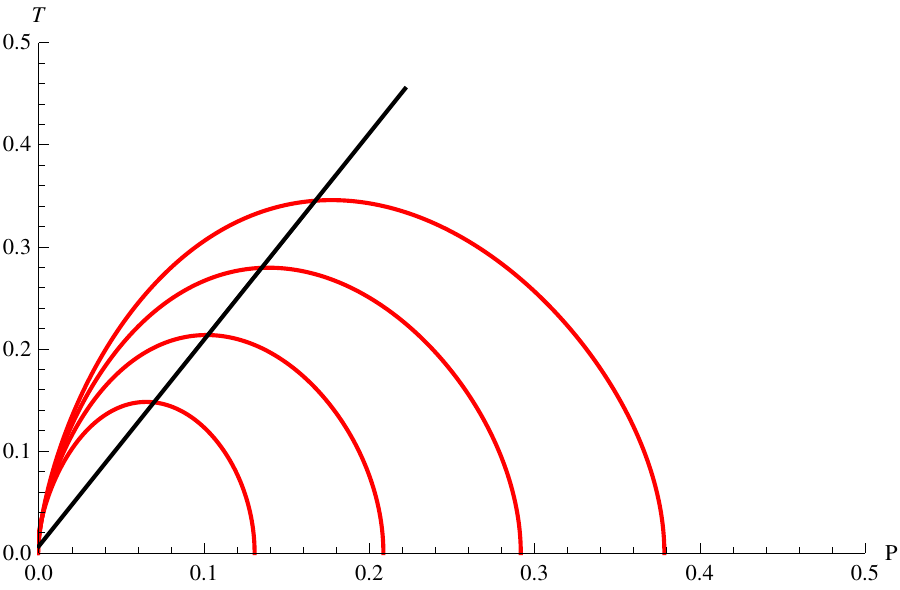}
\label{fig2.b}
\includegraphics[width=4cm,height=3cm]{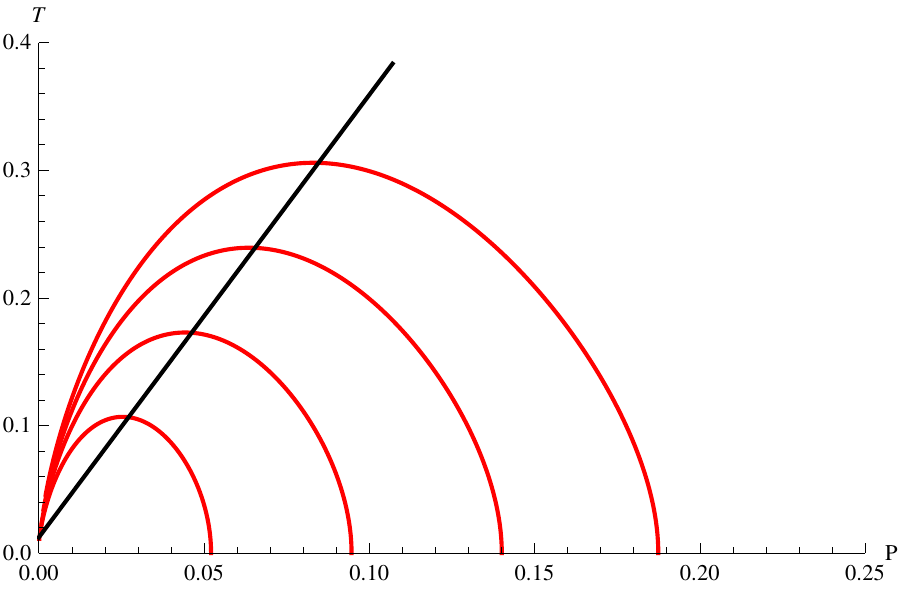}
\label{fig2.c}
\includegraphics[width=4cm,height=3cm]{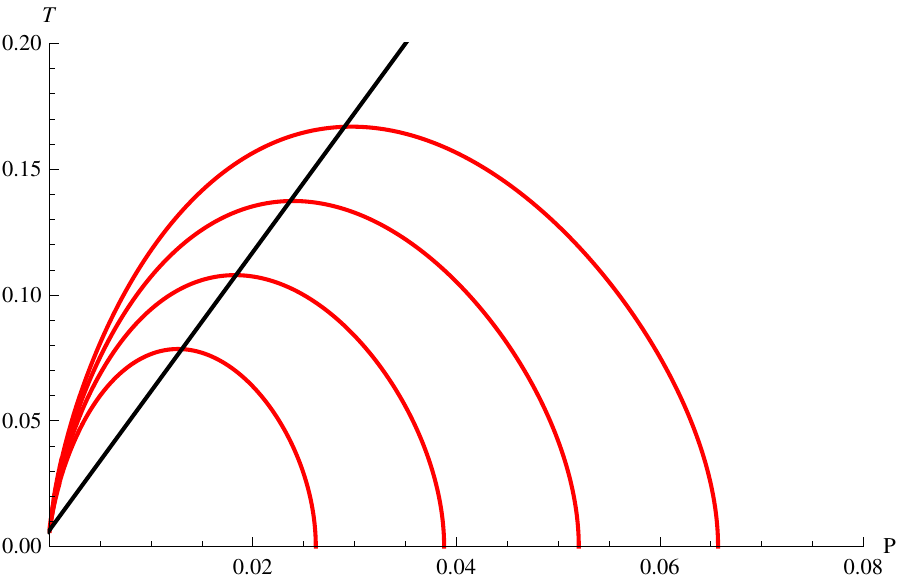}
\label{fig2.d}
\includegraphics[width=4cm,height=3cm]{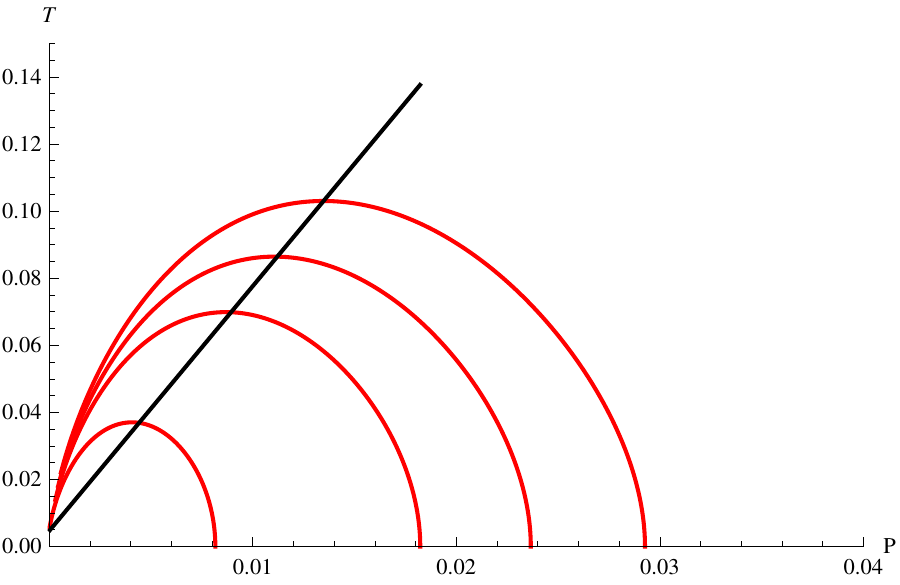}
\caption{Inversion and isenthalpic (constant mass) curves of Hayward-AdS black hole. Black and red lines present inversion and isenthalpic curves, respectively. From bottom to top, isenthalpic curves correspond to increasing values of mass $\mathcal{M}$ of theblack hole. (Top-left) $q=1$ and $M=1.5,2,2.5,3$. (Top-right) $q=2$ and $M=5,7,9,11$. (Bottom-left) $q=3$ and $M=8,10,12,14$. (Bottom-right) $q=4$ and $M=8,12,14,16$.}
\label{fig2}
\end{figure}
\par
In Fig.(\ref{fig1}), the inversion curves are presented for various values of magnetic charge $q$. In contrast to Van der Waals fluids, it can be seen from Fig.(\ref{fig1}) the inversion curves are not closed and there is only one inversion curve, as previously described in  \cite{35,36,37,38,39,40,41,42,43,44}. Obviously, with an increasing of $q$, the inversion temperature of the black hole for given pressure tends to increase.
\par
In Fig.(\ref{fig2}), we plot the isenthalpic (constant mass) and inversion curves for various values of magnetic charge $q$ in the $T-P$ plane. As it can be seen from Fig.(\ref{fig2}), the inversion curve divides the plane into two regions. The region above the inverse curve corresponds to the cooling region, while the region under the inversion curve corresponds to the heating region. Indeed, the heating and cooling regions are checked by the slope signs of the isenthalpic curve. The sign of slope is positive in the cooling region and it changes in the heating region. On the other hand, cooling (heating) does not happen on the inversion curve which plays role as a boundary between two regions.
%
%
%
\section{Conclusions}
\label{sec4}
%
%
%
\par\noindent
In this paper, we have investigated Joule-Thomson expansion for Hayward-AdS black hole in the extended phase space, where cosmological constant is viewed as pressure and the black hole mass as enthalpy. We have plotted the isenthalpic and inversion curves in the $T-P$ plane, and determine the cooling and heating regions for various values of magnetic charge $q$ and mass $\mathcal{M}$. It shown that, in contrast to Van de Waals fluids, the inversion curves are not closed and there is only one inversion curve. As a result, the isenthalpic curves have positive slope above the inversion curves so cooling occurs here, and the sign of slope changes under the inversion curves and heating occurs in this region.
%
\section{Acknowledgements}
\par\noindent
This work is supported by the Program for NCET-12-1060, by the Sichuan Youth Science and Technology Foundation with Grant No. 2011JQ0019, and by FANEDD with Grant No. 201319, and by Ten Thousand Talent Program of Sichuan Province, and by Sichuan Natural Science Foundation with Grant No. 16ZB0178, and by the starting funds of China West Normal University with Grant No.17YC513 and No.17C050.
%
%

%
%
\end{document}